


%
%


\def\famname{
 \textfont0=\textrm \scriptfont0=\scriptrm
 \scriptscriptfont0=\sscriptrm
 \textfont1=\textmi \scriptfont1=\scriptmi
 \scriptscriptfont1=\sscriptmi
 \textfont2=\textsy \scriptfont2=\scriptsy \scriptscriptfont2=\sscriptsy
 \textfont3=\textex \scriptfont3=\textex \scriptscriptfont3=\textex
 \textfont4=\textbf \scriptfont4=\scriptbf \scriptscriptfont4=\sscriptbf
 \skewchar\textmi='177 \skewchar\scriptmi='177
 \skewchar\sscriptmi='177
 \skewchar\textsy='60 \skewchar\scriptsy='60
 \skewchar\sscriptsy='60
 \def\rm{\fam0 \textrm} \def\bf{\fam4 \textbf}}
\def\sca#1{scaled\magstep#1} \def\scah{scaled\magstephalf} 
\def\twelvepoint{
 \font\textrm=cmr12 \font\scriptrm=cmr8 \font\sscriptrm=cmr6
 \font\textmi=cmmi12 \font\scriptmi=cmmi8 \font\sscriptmi=cmmi6 
 \font\textsy=cmsy10 \sca1 \font\scriptsy=cmsy8
 \font\sscriptsy=cmsy6
 \font\textex=cmex10 \sca1
 \font\textbf=cmbx12 \font\scriptbf=cmbx8 \font\sscriptbf=cmbx6
 \font\it=cmti12
 \font\sectfont=cmbx12 \sca1
 \font\refrm=cmr10 \scah \font\refit=cmti10 \scah
 \font\refbf=cmbx10 \scah
 \def\twelverm{\textrm} \def\twelveit{\it} \def\twelvebf{\textbf}
 \famname \textrm 
 \voffset=.04in \hoffset=.21in
 \normalbaselineskip=18pt plus 1pt \baselineskip=\normalbaselineskip
 \parindent=21pt
 \setbox\strutbox=\hbox{\vrule height10.5pt depth4pt width0pt}}


\catcode`@=11

{\catcode`\'=\active \def'{{}^\bgroup\prim@s}}

\def\screwcount{\alloc@0\count\countdef\insc@unt}   
\def\screwdimen{\alloc@1\dimen\dimendef\insc@unt} 
\def\screwbox{\alloc@4\box\chardef\insc@unt}

\catcode`@=12


\overfullrule=0pt			
\voffset=.04in \hoffset=.21in
\vsize=9in \hsize=6in
\parskip=\medskipamount	
\lineskip=0pt				
\normalbaselineskip=18pt plus 1pt \baselineskip=\normalbaselineskip
\abovedisplayskip=1.2em plus.3em minus.9em 
\belowdisplayskip=1.2em plus.3em minus.9em	
\abovedisplayshortskip=0em plus.3em	
\belowdisplayshortskip=.7em plus.3em minus.4em	
\parindent=21pt
\setbox\strutbox=\hbox{\vrule height10.5pt depth4pt width0pt}
\def\makefootline{\baselineskip=30pt \line{\the\footline}}
\footline={\ifnum\count0=1 \hfil \else\hss\twelverm\folio\hss \fi}
\pageno=1


\def\boxit#1{\leavevmode\thinspace\hbox{\vrule\vtop{\vbox{
	\hrule\kern1pt\hbox{\vphantom{\bf/}\thinspace{\bf#1}\thinspace}}
	\kern1pt\hrule}\vrule}\thinspace}
\def\Boxit#1{\noindent\vbox{\hrule\hbox{\vrule\kern3pt\vbox{
	\advance\hsize-7pt\vskip-\parskip\kern3pt\bf#1
	\hbox{\vrule height0pt depth\dp\strutbox width0pt}
	\kern3pt}\kern3pt\vrule}\hrule}}


\def\put(#1,#2)#3{\screwdimen\unit  \unit=1in
	\vbox to0pt{\kern-#2\unit\hbox{\kern#1\unit
	\vbox{#3}}\vss}\nointerlineskip}

%
%
%
%
%
%
%

\def\\{\hfil\break}

\def\center{\leftskip=0pt plus 1fill \rightskip=\leftskip \parindent=0pt
 \def\textindent##1{\par\hangindent21pt\footrm\noindent\hskip21pt
 \llap{##1\enspace}\ignorespaces}\par}
\def\unnarrower{\leftskip=0pt \rightskip=\leftskip}
\def\thetitle#1#2#3#4#5{
 \font\titlefont=cmbx12 \sca2 \font\footrm=cmr10 \font\footit=cmti10
  \twelverm
	{\hbox to\hsize{#4 \hfill ITP-SB-#3}}\par
	\vskip.8in minus.1in {\center\baselineskip=1.44\normalbaselineskip
 {\titlefont #1}\par}{\center\baselineskip=\normalbaselineskip
 \vskip.5in minus.2in #2
	\vskip1.4in minus1.2in {\twelvebf ABSTRACT}\par}
 \vskip.1in\par
 \narrower\par#5\par\unnarrower\vskip3.5in minus2.3in\eject}
\def\paper\par#1\par#2\par#3\par#4\par#5\par{\twelvepoint
	\thetitle{#1}{#2}{#3}{#4}{#5}} 
\def\author#1#2{#1 \vskip.1in {\twelveit #2}\vskip.1in}
\def\ITP{Institute for Theoretical Physics\\
	State University of New York, Stony Brook, NY 11794-3840}


\def\sect#1\par{\par\ifdim\lastskip<\medskipamount
	\bigskip\medskip\goodbreak\else\nobreak\fi
	\noindent{\sectfont{#1}}\par\nobreak\medskip} 
\def\itemize#1 {\item{[#1]}}	
\def\vol#1 {{\refbf#1} }		 

\def\ref#1{\setbox0=\hbox{M}$\vbox to\ht0{}^{#1}$}


\def\NP #1 {{\refit Nucl. Phys.} {\refbf B{#1}} }
\def\PL #1 {{\refit Phys. Lett.} {\refbf{#1}} }
\def\PR #1 {{\refit Phys. Rev. Lett.} {\refbf{#1}} }
\def\PRD #1 {{\refit Phys. Rev.} {\refbf D{#1}} }


\hyphenation{pre-print}
\hyphenation{quan-ti-za-tion}

%
%

\def\on#1#2{{\buildrel{\mkern2.5mu#1\mkern-2.5mu}\over{#2}}}
\def\dt#1{\on{\hbox{\bf .}}{#1}}                
\def\ddt#1{\on{\hbox{\bf .\kern-1pt.}}#1}    
\def\slap#1#2{\setbox0=\hbox{$#1{#2}$}
	#2\kern-\wd0{\hbox to\wd0{\hfil$#1{/}$\hfil}}}
\def\sla#1{\mathpalette\slap{#1}}                
\def\bop#1{\setbox0=\hbox{$#1M$}\mkern1.5mu
	\vbox{\hrule height0pt depth.04\ht0
	\hbox{\vrule width.04\ht0 height.9\ht0 \kern.9\ht0
	\vrule width.04\ht0}\hrule height.04\ht0}\mkern1.5mu}
\def\bo{{\mathpalette\bop{}}}                        
\def~{\widetilde} 
\mathcode`\*="702A                  
\def\in{\relax\ifmmode\mathchar"3232\else{\refit in\/}\fi} 
\def\f#1#2{{\textstyle{#1\over#2}}}	   
\def\half{{\textstyle{1\over{\raise.1ex\hbox{$\scriptstyle{2}$}}}}}

\catcode`\^^?=13				    
\catcode128=13 \def €{\"A}                 
\catcode129=13 \def {\AA}                 
\catcode130=13 \def '{\c}           	   
\catcode131=13 \def ƒ{\'E}                   
\catcode132=13 \def "{\~N}                   
\catcode133=13 \def …{\"O}                 
\catcode134=13 \def †{\"U}                  
\catcode135=13 \def ‡{\'a}                  
\catcode136=13 \def ˆ{\`a}                   
\catcode137=13 \def ‰{\^a}                 
\catcode138=13 \def Š{\"a}                 
\catcode139=13 \def ‹{\~a}                   
\catcode140=13 \def Œ{\alpha}            
\catcode141=13 \def {\chi}                
\catcode142=13 \def Ž{\'e}                   
\catcode143=13 \def {\`e}                    
\catcode144=13 \def {\^e}                  
\catcode145=13 \def '{\"e}                
\catcode146=13 \def '{\'\i}                 
\catcode147=13 \def "{\`\i}                  
\catcode148=13 \def "{\^\i}                
\catcode149=13 \def •{\"\i}                
\catcode150=13 \def –{\~n}                  
\catcode151=13 \def —{\'o}                 
\catcode152=13 \def ˜{\`o}                  
\catcode153=13 \def ™{\^o}                
\catcode154=13 \def š{\"o}                 
\catcode155=13 \def ›{\~o}                  
\catcode156=13 \def œ{\'u}                  
\catcode157=13 \def {\`u}                  
\catcode158=13 \def ž{\^u}                
\catcode159=13 \def Ÿ{\"u}                
\catcode160=13 \def  {\tau}               
\catcode161=13 \mathchardef ¡="2203     
\catcode162=13 \def ¢{\oplus}           
\catcode163=13 \def £{\relax\ifmmode\to\else\itemize\fi} 
\catcode164=13 \def ¤{\subset}	  
\catcode165=13 \def ¥{\infty}           
\catcode166=13 \def ¦{\mp}                
\catcode167=13 \def §{\sigma}           
\catcode168=13 \def ¨{\rho}               
\catcode169=13 \def ©{\gamma}         
\catcode170=13 \def ª{\leftrightarrow} 
\catcode171=13 \def «{\relax\ifmmode\acute\else\expandafter\'\fi}
\catcode172=13 \def ¬{\relax\ifmmode\expandafter\ddt\else\expandafter\"\fi}
\catcode173=13 \def ­{\equiv}            
\catcode174=13 \def ®{\approx}          
\catcode175=13 \def ¯{\Omega}          
\catcode176=13 \def °{\otimes}          
\catcode177=13 \def ±{\ne}                 
\catcode178=13 \def ²{\le}                   
\catcode179=13 \def ³{\ge}                  
\catcode180=13 \def ´{\upsilon}          
\catcode181=13 \def µ{\mu}                
\catcode182=13 \def ¶{\delta}             
\catcode183=13 \def ·{\epsilon}          
\catcode184=13 \def ¸{\Pi}                  
\catcode185=13 \def ¹{\pi}                  
\catcode186=13 \def º{\beta}               
\catcode187=13 \def »{\partial}           
\catcode188=13 \def ¼{\nobreak\ }       
\catcode189=13 \def ½{\zeta}               
\catcode190=13 \def ¾{\sim}                 
\catcode191=13 \def ¿{\omega}           
\catcode192=13 \def À{\dt}                     
\catcode193=13 \def Á{\gets}                
\catcode194=13 \def Â{\lambda}           
\catcode195=13 \def Ã{\nu}                   
\catcode196=13 \def Ä{\phi}                  
\catcode197=13 \def Å{\xi}                     
\catcode198=13 \def Æ{\psi}                  
\catcode199=13 \def Ç{\int}                    
\catcode200=13 \def È{\oint}                 
\catcode201=13 \def É{\relax\ifmmode\cdot\else\vol\fi}    
\catcode202=13 \def Ê{\relax\ifmmode\,\else\thinspace\fi}
\catcode203=13 \def Ë{\`A}                      
\catcode204=13 \def Ì{\~A}                      
\catcode205=13 \def Í{\~O}                      
\catcode206=13 \def Î{\Theta}              
\catcode207=13 \def Ï{\theta}               
\catcode208=13 \def Ð{\relax\ifmmode\bar\else\expandafter\=\fi}
\catcode209=13 \def Ñ{\overline}             
\catcode210=13 \def Ò{\langle}               
\catcode211=13 \def Ó{\relax\ifmmode\{\else\ital\fi}      
\catcode212=13 \def Ô{\rangle}               
\catcode213=13 \def Õ{\}}                        
\catcode214=13 \def Ö{\sla}                      
\catcode215=13 \def ×{\relax\ifmmode\check\else\expandafter\v\fi}
\catcode216=13 \def Ø{\"y}                     
\catcode217=13 \def Ù{\"Y}  		    
\catcode218=13 \def Ú{\Leftarrow}       
\catcode219=13 \def Û{\Leftrightarrow}       
\catcode220=13 \def Ü{\relax\ifmmode\Rightarrow\else\sect\fi}
\catcode221=13 \def Ý{\sum}                  
\catcode222=13 \def Þ{\prod}                 
\catcode223=13 \def ß{\widehat}              
\catcode224=13 \def à{\pm}                     
\catcode225=13 \def á{\nabla}                
\catcode226=13 \def â{\quad}                 
\catcode227=13 \def ã{\in}               	
\catcode228=13 \def ä{\star}      	      
\catcode229=13 \def å{\sqrt}                   
\catcode230=13 \def æ{\^E}			
\catcode231=13 \def ç{\Upsilon}              
\catcode232=13 \def è{\"E}    	   	 
\catcode233=13 \def é{\`E}               	  
\catcode234=13 \def ê{\Sigma}                
\catcode235=13 \def ë{\Delta}                 
\catcode236=13 \def ì{\Phi}                     
\catcode237=13 \def í{\`I}        		   
\catcode238=13 \def î{\iota}        	     
\catcode239=13 \def ï{\Psi}                     
\catcode240=13 \def ð{\times}                  
\catcode241=13 \def ñ{\Lambda}             
\catcode242=13 \def ò{\cdots}                
\catcode243=13 \def ó{\^U}			
\catcode244=13 \def ô{\`U}    	              
\catcode245=13 \def õ{\bo}                       
\catcode246=13 \def ö{\relax\ifmmode\hat\else\expandafter\^\fi}
\catcode247=13 \def÷{\relax\ifmmode\tilde\else\expandafter\~\fi}
\catcode248=13 \def ø{\ll}                         
\catcode249=13 \def ù{\gg}                       
\catcode250=13 \def ú{\eta}                      
\catcode251=13 \def û{\kappa}                  
\catcode252=13 \def ü{\half}     		 
\catcode253=13 \def ý{\Gamma} 		
\catcode254=13 \def þ{\Xi}   			
\catcode255=13 \def ÿ{\relax\ifmmode{}^{\dagger}{}\else\dag\fi}


\def\ital#1Õ{{\it#1\/}}	     
\def\un#1{\relax\ifmmode\underline#1\else $\underline{\hbox{#1}}$
	\relax\fi}

\def\tdt#1{\on{\hbox{\bf .\kern-1pt.\kern-1pt.}}#1}   
\def\({\eqno(}
\def\li{\eqalignno}
\def\refs{\sect{REFERENCES}\par\medskip \frenchspacing 
	\parskip=0pt \refrm \baselineskip=1.23em plus 1pt
	\def\ital##1Õ{{\refit##1\/}}}


\def\õ#1{
	\screwcount\num
	\num=1
	\screwdimen\downsy
	\downsy=-1.5ex
	\mkern-3.5mu
	õ
	\loop
	\ifnum\num<#1
	\llap{\raise\num\downsy\hbox{$õ$}}
	\advance\num by1
	\repeat}
\def\upõ#1#2{\screwcount\numup
	\numup=#1
	\advance\numup by-1
	\screwdimen\upsy
	\upsy=.75ex
	\mkern3.5mu
	\raise\numup\upsy\hbox{$#2$}}


\catcode`\|=\active \catcode`\<=\active \catcode`\>=\active 
\def|{\relax\ifmmode\delimiter"026A30C \else$\mathchar"026A$\fi}
\def<{\relax\ifmmode\mathchar"313C \else$\mathchar"313C$\fi}
\def>{\relax\ifmmode\mathchar"313E \else$\mathchar"313E$\fi}


\def\Ä{\varphi}	\def\¢{\ominus}

\paper

SIMPLIFYING ALGEBRA\\ IN FEYNMAN GRAPHS\\
Part II: Spinor Helicity from the Spacecone

\author{G. Chalmers and W. Siegel\footnote{${}^1$}{
 Internet addresses: chalmers and siegel@insti.physics.sunysb.edu.}}\ITP

98-05

January 5, 1997

Manifestly Lorentz covariant Feynman rules are given in terms of a ``scalar"
field for each helicity, dramatically simplifying the calculation of amplitudes
with massless particles.  The spinor helicity formalism is properly identified
as a null complex spacelike (not lightlike) gauge, where two  massless
external momenta define the reference frame.  Usually this gauge is applied
only to external line factors; we extend this method to vertices and
propagators by modifying the action itself using lightcone methods.  

Ü1. Introduction

In other papers in this series [1] we describe a method to simplify
Feynman diagram calculations for massive fields, based on the observation
that fields with only undotted spinor indices formally have the same structure
as nonrelativistic fields, since a Lorentz transformation on such fields is just a
complex rotation.  Thus spin $s$ is described by a field with $2s+1$
components in a manifestly covariant way: spin 1/2 by a chiral spinor, spin 1
by a self-dual tensor, without their complex conjugates.  The result is a
unitary gauge where all propagators are simply $1/(p^2+m^2)$.  Also, external
line factors are unconstrained (i.e., arbitrary), so amplitudes can be
completely evaluated covariantly (rather than leaving covariance for the
more numerous squares of terms in probabilities).  Many supersymmetry
relations are already obvious from the graphs, since actions for spins 1/2 and
1 more closely resemble those for spin 0.  Much repetitive algebra ordinarily
applied in graphs is already done once and for all in the action itself (so the
action has more vertices, such as seagulls for spinors,
but the amplitude has fewer terms).  Since the fields are chiral/self-dual, this
method is particularly suited to graphs with maximal helicity violation (MHV),
which are those that have the simplest final form.

Although this method can be applied to massless theories by an appropriate
limiting procedure (which is trivial for spin 1/2), it is inherently tailored to
massive theories.  In particular, the concept of using only the physical
degrees of freedom suggests the further reduction from $2s+1$ components
for the massive case to just 2 for the massless case.  In this paper we
accomplish this task by combining two well-known methods, the lightcone
formalism [2] and twistors [3] (the spinor helicity formalism [4] for external
line factors).

The lightcone formalism is defined by choosing a fixed lightlike direction
(``$-$"), choosing the gauge where the corresponding lightlike components of
gauge fields vanish, and eliminating all auxiliary degrees of freedom (``$+$")
by their nonpropagating (no ``$+$" derivatives) equations of motion.  This last
step is the most important one:  It reduces all nonzero spins to two degrees
of freedom (corresponding to the two helicities); using 4-vectors to describe
lightcone Yang-Mills would miss the whole point of the lightcone approach. 
This gauge is manifestly unitary, and all propagators are $1/p^2$ because
each field has a single component (complex to describe two helicities).  The
main disadvantage of the lightcone is that Lorentz invariance is not manifest: 
Although the action is Lorentz invariant, Lorentz transformations are
nonlinear in the fields (and momenta), and the Feynman rules are asymmetric
in the various components of the momenta.  An especially simple case is
self-dual Yang-Mills theory [5], whose Feynman rules [6] are manifestly
covariant under half the Lorentz group (in 2+2 dimensions, one of the two
SL(2)'s of SO(2,2) = SL(2)$°$SL(2)).  In particular, the self-dual rules are much
simpler than the usual ones for deriving MHV amplitudes.

The spinor helicity method translates all external line factors into twistor
notation, essentially using the Penrose transform for free fields.  Each
massless momentum is expressed as the square of a momentum twistor, while
the external line factor (i.e., the external free field) is expressed in terms of
that and (for gauge fields) an arbitrary gauge-dependent polarization twistor. 
If the polarization twistors for all external lines are chosen to point in the
same fixed direction, then the gauge is a lightcone gauge, but only for the
external fields (and not the internal lines).  However, in practice one chooses
to equate the polarization twistors with some of the momentum twistors of
the other lines, so the amplitude is expressed completely in terms of
momentum twistors (and the probability completely in terms of momenta) 
[4]. 
Unlike the lightcone formalism, with this approach not only the vertices but
even the external line factors are manifestly covariant.  

On the other hand, spinor helicity modifies only the external line factors, and
does not simplify the vertices.  The result is that, although the final
expression for MHV amplitudes is extremely simple [7], their calculation does
not fully reflect this simplicity.  Explicitly, the first two steps in a spinor
helicity calculation are:  (1) Write down all the terms that come from the
various graphs, using the usual Feynman rules in color-ordered form [4]. 
Some simplification comes from using the Gervais-Neveu gauge, which gives
only 3 terms for each 3-point vertex and 1 term for each 4-point (for pure
Yang-Mills theory).  (2) Contract all the vector indices (and spinor ones for
quarks), coming from vertices, propagator numerators, and external line
factors.  A remarkable feature of spinor helicity is the large number of terms
that vanish in the second step:  Many vanish immediately, from the direct
effect the external line factors have on the vertices; but most vanish only
after more careful inspection, when polarization vectors are contracted from
different parts of the graph via the Kronecker deltas in the propagators.  An
equally remarkable feature of spinor helicity is that one has to spend a
considerable fraction of the calculation writing down these terms that don't
contribute.  Generally, any method that requires calculation to produce zeros
indicates a simpler method is missing for which the zeros would be automatic
(e.g., supergraphs for the ``miraculous cancellations" of supersymmetry).

In this paper we derive new Feynman rules that avoid the second step
above altogether by eliminating Lorentz indices on all massless fields, while
retaining all the advantages of spinor helicity.  This is accomplished by
modifying the usual lightcone method to incorporate the principles of the
spinor helicity method:  (1) Although we choose arbitrary fixed vectors to
define the reference frame used to derive the lightcone action, in the
amplitudes we choose to identify the two lightlike axes with two of the
physical external (``reference") momenta (as allowed by Lorentz and gauge
invariance), which also defines the other two (spacelike) axes, using
momentum twistors.  The result is then manifestly covariant, since explicit
Lorentz components are replaced with Lorentz invariants.  On external lines,
this is equivalent to spinor helicity, but we have extended the method to
internal lines.  (2) As a result, the direction chosen to define the gauge is not
lightlike, but spacelike and complex, although it is still null.  The gauge
condition is thus complex, like the Gervais-Neveu gauge, although it is linear. 
(This distinction does not exist in $2+2$ dimensions, but the Wick rotation to
3+1 is different.)  We thus refer to our formalism as the ``spacecone"
(although, as for ``lightcone", no cone is actually involved; it is an
abbreviation for ``complex spacelike null hyperplane").  (3) In the usual
lightcone approach, where a fixed component of the gauge field vanishes, one
takes care to avoid the vanishing of the corresponding lightlike component of
the momentum, which would produce singularities.  In our approach its
vanishing is ÓrequiredÕ by definition on the two reference lines.  We show that
the external line factors cancel such singularities, and find the resulting
vertices for those two lines are much simpler than the rest.  This accounts for
much of the simplification of the spinor helicity formalism.  These external
line factors follow from the covariant ones used in the usual spinor helicity
approach, and are not the trivial ones normally used in the lightcone
approach, although they are still simple.  (The field redefinition that relates
the two is singular for the reference lines.)  

The main advantage of our approach over the usual spinor helicity is that
there are no Lorentz indices associated with any lines, although they have an
orientation (+ helicity at one end, $-$ at the other); all indices show up at the
final step, when Lorentz invariants are expressed in terms of momentum
twistors.  In particular, this means that vanishing graphs and terms in graphs,
which in the spinor helicity approach would be seen by expanding the vertices
and performing the (vector and spinor) index algebra, are avoided in the
spacecone approach without performing any algebra whatsoever.

In the following section we review the basic points of the lightcone
formalism, including special features of four dimensions, such as
simplification of the interactions and self-duality.  In the next section we
review twistors and spinor helicity.  The spacecone is introduced in section 4. 
Examples are given in section 5 to illustrate the improvement over previous
techniques.  Recursion relations are discussed in section 6.  The final section
contains our conclusions.

Ü2. Lightning review of lightcone

We start with the Lagrangian for Yang-Mills theory, in a convenient
normalization ($S=tr¼L/g^2$),
$$ L = \f18 F^2,ââF^{ab} = »^{[a}A^{b]} +i[A^a,A^b]  
\(1) 
$$
 where $»^{[a}A^{b]}=»^a A^b-»^b A^a$, and the indices $a,b$ are 
vector indices.  (Throughout this paper we will use pure Yang-Mills theory as
our standard example, with straightforward generalization to other spins.) 
We use upper-case letters to denote vectors and lower-case to denote their
(contravariant) lightcone components, as
$$ A^a = (a,Ða,a^+,a^-)  
\(2) 
$$
 In Minkowski space $Ða$ is the complex conjugate of $a$, while $a^à$ are
real.  Defining our lightcone frame by the Minkowksi-space inner product
$$ AÉB = aÐb +Ðab -a^+ b^- -a^- b^+  
\(3) 
$$
we choose the lightcone gauge
$$ a^- = 0 
\(4) 
$$
 (In the original description of the lightcone gauge, the formalism was
described by an ``infinite-momentum frame".  Of course, Lorentz invariance
means the equations are frame-independent, and shortly thereafter it was
realized that the infinite-momentum limit was a misconception, and could be
replaced with a simple change of variables.  Even the parton model was
originally described in this language, until it was realized that it was only
hiding the Lorentz-invariant statement that the one physical assumption of
the parton model is a transverse momentum cutoff, which is realized
dynamically in QCD via asymptotic freedom.)

The equation of motion for $a^+$ now contains no ``time" derivatives $»^+$,
so we use this equation to eliminate this ``auxiliary" component from the
Lagrangian:  After a little algebra, we find
$$ L = Ða»^+ »^- a +\f14(F^{-+})^2 -\f14(F^{tÐt})^2 
\(5) 
$$
$$ F^{-+} = »Ða +лa +i{1\over »^-}([a,»^- Ða] +[Ða,»^- a]),ââ
	F^{tÐt} = »Ða -лa +i[a,Ða] 
\(6) 
$$
 Only in four dimensions can we simplify the Lagrangian by using the self-dual
and anti-self-dual combinations (which would have been indicated if we had
used spinor notation):
$$ 
\li{ {\cal F} & = ü(F^{-+}+F^{tÐt}) = »Ða +i{1\over »^-}[a,»^- Ða],ââ
		Ð{\cal F} = ü(F^{-+}-F^{tÐt}) = лa +i{1\over »^-}[Ða,»^- a] \cr
	L & = Ða»^+ »^- a +{\cal F}Ð{\cal F}  \cr
	& = -üÐaõa -i\left({л\over »^-}a\right)[a,»^-Ða]
		-i\left({»\over »^-}Ða\right)[Ða,»^-a]
		+[a,»^-Ða]{1\over (»^-)^2}[Ða,»^-a] \cr
	üõ & = »Ð»  - »^+ »^- &(7) \cr}  
$$
The Feynman rules may be obtained in a straigtforward fashion;   
for example, the color-ordered three-point vertex contains only 
two terms and is thus simpler than usual gauges, such as 
Fermi-Feynman or Gervais-Neveu.  

Translation into van der Waerden notation for Weyl spinors is easy:  In terms
of our lightcone basis, we have the Hermitian 2$ð$2 matrix
$$ A^{ŒÀº} = \pmatrix{ a^+ & Ða \cr a & a^- \cr},ââA^2 = -2¼det¼A 
\(8)
$$
 Spinor indices are raised and lowered with the antisymmetric symbol
$$ C_{Œº} = C_{ÀŒÀº} = -C^{Œº} = -C^{ÀŒÀº}
	 = {\textstyle\left({0\atop i}{-i\atop 0}\right)} 
\(9) 
$$
$$ Æ^Œ = C^{Œº}Æ_º,âÆ_Œ = Æ^º C_{ºŒ},âÐÆ^{ÀŒ} = (Æ^Œ)ÿ = C^{ÀŒÀº}ÐÆ_{Àº} 
\(10) 
$$
$$ AÉB = A^{ŒÀº}B_{ŒÀº} 
\(11) 
$$
 Note that each of the 4 lightcone/spinor-notation components is defined with
respect to a null direction:  Two are real and lightlike ($ötàöx$), two are complex
and spacelike ($öyàiöz$).  (In 2+2 dimensions all 4 are real and lightlike, while in
4+0 dimensions all are complex and spacelike.)

Spinors can be treated similarly:  There is no gauge condition for spin 1/2, but
half of the field is auxiliary.  Now spinor notation is necessary, and in the
lightcone a Weyl spinor reduces to a single complex component, as did the
vector.

Self-dual Yang-Mills theory also is described more easily in spinor notation,
where the field strength naturally separates into self-dual and anti-self-dual
parts:
$$ [á^{ŒÀº},á^{©À¶}] = i(C^{Œ©}f^{ÀºÀ¶} +C^{ÀºÀ¶}f^{Œ©}) 
\(12) 
$$
 Self-duality $f^{μ}=0$ is then partially solved in the lightcone gauge by
$$ f^{--} = 0âÜâ[á^{-Àº},á^{-À¶}] = 0âÜâA^{-Àº} = 0â(a^- = a = 0)  
\(13) 
$$
$$ f^{-+} = 0âÜâ»^{-[Àº}A^{+À¶]} = 0âÜâA^{+Àº} = »^{-Àº}Äâ
	(a^+ = »Ä,¼Ða = »^- Ä)  
\(14) 
$$
The remaining equation $f^{++}=0$ is identical to that found from the above
lightcone action by varying with respect to $a$ and then setting $a=0$:  It is
the ordinary Yang-Mills field equation restricted to one helicity.  If we define
self-dual Yang-Mills by the Lagrangian $G_{μ}f^{μ}$, then our lightcone
analysis leads directly to the first two terms of the above lightcone action,
with $(»^-)^{-1}G_{++}$ acting as the replacement for $a$.

Ü3. Twistors and spinor helicity

The basic principle of the twistor approach is that a lightlike 4-vector can be
written as the square of a commuting spinor (``twistor"):
$$ P^2 = 0âÜâP^{ŒÀº} = àp^Œ Ðp^{Àº} 
\(15) 
$$
 where the sign depends on whether the vector is forward or backward (with
respect to the time direction):  In practice we ignore this sign, as justified by
crossing symmetry, or by Wick rotation from 2+2 dimensions, where $p^Œ$
and $Ðp^{ÀŒ}$ are independent and real (rather than complex conjugates), so
the sign can be absorbed into either.  This representation is particularly useful
for free, massless particles:  Using this twistor to describe the momentum, the
usual free field equations can be solved for the free field strengths as
$$ f^{Œ_1...Œ_{2h}} = p^{Œ_1}òp^{Œ_{2h}}\Ä_h 
\(16) 
$$
for a spinor $f^Œ$, vector field strength $f^{Œº}$, Weyl tensor $f^{Œº©¶}$,
etc.  Thus, massless degrees of freedom are immediately reduced to
(complex) scalars.  Actually, Lorentz transformations are reduced to their
little group, helicity, which transforms the twistor by a phase (independently
of the usual Lorentz), and thus the ``scalar" by the helicity $h$ times the
phase.

An old lesson learned from supersymmetry is that all spinor algebra in four
dimensions is performed most conveniently with two-component spinor
indices exclusively, and this rule also gives the simplest calculations in the
spinor helicity formalism.  This is much simpler than algebra with
four-component (Dirac) spinors.  In particular, Fierz identities are avoided. 
This also means one should avoid all $©$ and $§$ matrices, since they are
nothing more than Clebsch-Gordan coefficients.  Vectors are described only as
objects with two two-component indices.  

Since almost all spinor algebra for spins $²1$ involves objects carrying at most
two spinor indices (spinors, vectors, self-dual tensors), for such purposes it is
usually convenient to use matrix notation, defined by
$$ Òp| = p^Œ,â|pÔ = p_Œ;â
	[p| = p^{ÀŒ},â|p] = p_{ÀŒ} 
\(17) 
$$
 As a result, we also have
$$ ÒpqÔ = p^Œ q_Œ,â[pq] = p^{ÀŒ}q_{ÀŒ};âÒpqÔ* = [qp] = -[pq]  
\(18) 
$$
$$ P = P_Œ{}^{Àº},âP* = -P^º{}_{ÀŒ},âÒk|P|q] = k^Œ P_Œ{}^{Àº} q_{Àº},â
	PK* +KP* =(PÉK)I  
\(19) 
$$
$$ f = f_Œ{}^º,âf* = f_{ÀŒ}{}^{Àº},âÒp|f|qÔ = p^Œ f_Œ{}^º q_º 
\(20) 
$$
$$ ÒpqÔÒrsÔ +ÒqrÔÒpsÔ +ÒrpÔÒqsÔ = 0 
\(21) 
$$
 where the last (``Schouten") identity is the result of antisymmetrizing 3
indices that take only 2 values.  (Note that $ÒpqÔ=-ÒqpÔ$, $[pq]=-[qp]$ are
consequences of using twistors; when using physical, anticommuting spinors,
$ÒƍÔ=+ҍÆÔ$, and $ÒÆÆÔ±0$ occurs in mass terms.)

Although gauge-invariant objects can be written directly in terms of
momentum twistors and scalars, the same is not true for gauge fields.  Thus
for a gauge vector we write (in matrix notation)
$$ A = {|·Ô[p|\over Ò·pÔ}Ð\ÄâÜâf* = |p][p|Ð\Ä,âf = 0 
\(22) 
$$
 for + helicity or
$$ A = {|pÔ[·|\over [·p]}\ÄâÜâf = |pÔÒp|\Ä,âf* = 0 
\(23) 
$$
 for $-$.  (Positive helicity is the same as self-duality, negative is
anti-self-dual.)  The former agrees with the self-dual Yang-Mills result of the
previous section, up to normalization, if we choose
$$ ·^Œ ¾ ¶^Œ_+ 
\(24) 
$$
 These expressions are used for external lines in Feynman diagrams:  Setting
scalar fields $\Ä=1$ as usual, the factors multiplying $\Ä$ and $Ð\Ä$ are
identified as the external line factors for a massless vector.  Thus, the
twistor formalism is essentially a covariant way of writing the axial
gauge
$$ NÉA = 0,âN = |·Ô[·| 
\(25) 
$$
in terms of an arbitrary lightlike vector $N$.

However, a major simplification is achieved in applying spinor helicity
methods to explicit evaluation of Feynman diagrams:  Instead of choosing the
lightlike vectors $N$ to be the same on each external line, as in a lightcone
gauge, they are chosen to vary from line to line (i.e., to be momentum
dependent).  Furthermore, these lightlike vectors, rather than being arbitrary,
are identified with some of the external lightlike (massless) momenta
(although, of course, no $N$ is identified with the momentum of the same
line).  One result is that all amplitudes are manifestly Lorentz invariant, since
they are expressed completely in terms of invariant products of momentum
twistors (and no other Lorentz structures).

Ü4. Spacecone

Although these principles should be enough to appreciate the advantages of
spinor helicity methods, there is one additional rule that, although not
required, always gives the simplest results in practice:  All the $N$'s are
chosen the same for one sign of the helicity, and all the same for the other
sign.  This means that all the $·$'s (+ helicity) are the same undotted spinor
$|+Ô$, and all the $з$'s ($-$ helicity) are the same dotted spinor $|-]$.  However,
because no $N$ can be identified with a $P$ of the same line, this means that
the $·$'s are identified with a $P=|+Ô[+|$ of a line with the opposite sign
helicity ($-$), and all $з$'s with a $P=|-Ô[-|$ of a line with + helicity.  In
particular, this means that the one $з$ used on all such lines is not the
complex conjugate of the one $·$ used on all the other lines.  The naive way
to interpret this would be to say that all the lines of positive helicity are in
one lightcone gauge, and all the lines of negative helicity are in a different
lightcone gauge.  However, a much simpler way to interpret this is to say that
ÓallÕ lines are in the same gauge, defined by the vector
$$ N = |+Ô[-|,âNÉA = 0 
\(26) $$
 This $N$ is complex; it is also spacelike, since $ÒppÔ=0$, so it is orthogonal to
the two lightlike vectors (external momenta) $|àÔ[à|$, as well as being null. 
Thus, we have a complex spacelike (axial) gauge.

Repeating the derivation of the lightcone action with the gauge
$$ a = 0 
\(27) $$
 eliminating $Ða$ by its equation of motion, we now have:
$$ L = ü a^+õa^- -i\left({»^-\over »}a^+\right)[a^+,»a^-]
		-i\left({»^+\over »}a^-\right)[a^-,»a^+]
		+[a^+,»a^-]{1\over »^2}[a^-,»a^+] 
\(28) 
$$
 The fact that the three vertices come in helicities ++$-$, $--$+, and ++$--$
will prove important later when writing Feynman graphs, since now the $à$
index on the field labels its helicity (as seen, e.g., from the external line
factors of the previous section with our present choice of $·$ and $з$).  This
action is complex because of the appearance of $»$ without $л$.

In spinor notation in matrix form, we use the basis $|+Ô[+|$, $|-Ô[-|$, $|-Ô[+|$,
$|+Ô[-|$:
$$ P = p^+ |+Ô[+| +p^- |-Ô[-| +p|-Ô[+| +Ðp|+Ô[-| 
\(29) 
$$
 with the normalization
$$ Ò+-Ô = [-+] = 1 
\(30) $$
 so we can write, e.g., for massless momentum $P=|pÔ[p|$,
$$ p^+ = Òp-Ô[-p],âp^- = Ò+pÔ[p+],âp = Ò+pÔ[-p],âÐp = Òp-Ô[p+] 
\(31) $$
 (We can then identify $à$ as labelling rows and columns in 2$ð$2 matrix
notation; with our normalization, $Ò+|=¶_+^Œ$ but $Ò-|=-i¶_-^Œ$, so this leads
to the harmless redefinition  $A^{ŒÀº}=({a^+\atop -ia}¼{iÐa\atop a^-})$.)

This is as much as we can do in the action; however, when we write down an
explicit amplitude, we identify $|+Ô[+|$ with the momentum of one line with
negative helicity (vector or, more generally, spinor), and $|-Ô[-|$ with that of a
line of positive helicity.  (Alternatively, we can take one functional derivative
each with respect to $a^+$ and $a^-$ of the S-matrix generating functional,
to get a propagator in a background, and define the functional integral in
terms of the momenta associated with the ends of the propagator.  Then the
fields in the action in this functional integral become true scalars.)  This
defines $|àÔ$ up to phases, and thus $|-Ô[+|$ and $|+Ô[-|$, on which the phase
transformation is a rotation in the plane orthogonal to the two momenta. 
Thus our choice of the phase $Ò+-Ô/[-+]=1$ is a further
specification of these phases, while our choice of the magnitude
$Ò+-Ô[-+]=-Ò+|[+|É|-Ô|-]=1$ is a choice of (mass) units.  In explicit calculations,
we restore generality (in particular, to allow momentum integration) by
inserting appropriate powers of $Ò+-Ô$ and $[-+]$ at the end of the
calculations, as determined by simple dimensional and helicity analysis.  (This
avoids a clutter of normalization factors $å{Ò+-Ô[-+]}$ at intermediate
stages.)  For example, looking at the form of the usual spinor helicity external
line factors, and counting momenta in the usual Feynman rules, we see that
any tree amplitude (or individual graph) in pure Yang-Mills must go as
$$ Ò¼Ô^{2-E_+}[¼]^{2-E_-} 
\(32) $$
 where $E_à$ is the number of external lines with helicity $à$.

We now return to external line factors.  The naive factors for the above
Lagrangian are 1, since the kinetic term resembles that of a scalar.  However,
this would lead to unusual normalization factors in probabilities, which are
not obvious in this complex gauge.  Therefore, we determine external line
factors from the earlier spinor helicity expressions for external 4-vectors.  
$$ (·_+)^+ = -Ò-|[-|É{|+Ô|p]\over Ò+pÔ} = {[-p]\over Ò+pÔ} 
\(33) $$
$$ (·_-)^- = -Ò+|[+|É{|pÔ|-]\over [-p]} = {Ò+pÔ\over [-p]} 
\(34) $$
 Note that these factors are inverses of each other, consistent with leaving
invariant (the inner product defined by) the kinetic term.

An exception is the external line factors for the reference momenta
themselves, where $|pÔ=|¦Ô$ for helicity $à$ gives vanishing results.  However,
examination of the Lagrangian shows this zero can be canceled by by a $1/»$
in a vertex, since $p=Ðp=0$ for the reference momenta by definition.  (Such
cancellations occur automatically from field redefinitions in the lightcone
formulation of the self-dual theory.)  The actual expressions we want to
evaluate, before choosing the reference lines, are then
$$ {p^-\over p}(·_+)^+ = {Ò+pÔ[p+]\over Ò+pÔ[-p]}¼{[-p]\over Ò+pÔ} = 
	{[p+]\over Ò+pÔ} \(35) $$
$$ {p^+\over p}(·_-)^- = {Òp-Ô[-p]\over Ò+pÔ[-p]}¼{Ò+pÔ\over [-p]} =
	{Òp-Ô\over [-p]} \(36) $$
 Evaluating the former at $|pÔ=|-Ô$ and the latter at $|pÔ=|+Ô$, we get 1 in both
cases.  In summary, for reference lines: (1) use only the 3-point vertex of the
corresponding self-duality ($àà¦$ for helicity $à$), and use only the term
associating the singular factor with the reference line (the other term and the
other vertices give vanishing contributions); (2) including the momentum
factors on that line from the vertex, the external line factor is 1.

Obviously, the uniqueness of the vertex term for a reference line is a
considerable additional simplification for the rules.  As examples, consider the
amplitudes in pure Yang-Mills that are known to vanish by supersymmetry
[8]:  By simple counting of +'s and $-$'s, we see that the tree graphs with the
fewest external $-$'s, those with only self-dual vertices (++$-$), have a single
external $-$.  Thus the all + amplitude vanishes automatically.  Furthermore,
the diagrams with a single external $-$ must have that line chosen as one of
the reference lines.  However, by the above rules that line can carry only the
ÓantiÕ-self-dual vertex ($--$+), so those amplitudes also vanish.

The vanishing of these amplitudes is also easy to see in the usual spinor
helicity formalism:  In the spacecone gauge, the only nonvanishing inner
products of polarization vectors are between those of opposite helicity,
neither of which can be a reference line.  Since by momentum counting a tree
graph in pure Yang-Mills theory must have at least one such product (as
opposed to momentum times polarization), nonvanishing graphs must have at
least two external lines of each helicity (including reference lines) [4].  The
amazing thing about our spacecone approach is that our corresponding
arguments of the previous paragraph make no reference to momenta or inner
products; this extends to the avoidance of vanishing terms in nonvanishing
amplitudes, where the usual spinor helicity methods require explicit index
algebra.

The vanishing amplitudes can also be seen from supersymmetry.  On the other
hand, supersymmetry is not required on the spacecone, because it is already
built in:  The half of the supersymmetry transformations actually used are
trivial on the spacecone (or lightcone), reflecting the fact that the terms in
the Lagrangian are identical for different helicities except for the placement
of factors of the transverse momentum component $p$ (or, in the usual
lightcone formalism, the longitudinal component $p^+$).  In fact, in the
self-dual theory, because of field redefinitions, the terms are identical
without exception.

Ü5. More examples

The simplest nonvanishing amplitude is ++$--$.  We use color ordering; i.e., we
examine only planar diagrams for each permutation of external lines.  We
consider the case where the helicities are cyclically ordered as ++$--$; we
label them 1234, and choose 1 and 4 as the reference lines; this amplitude
can be denoted as $¢$+$-\¢$.  ($P_4=|+Ô[+]$, $P_1=|-Ô[-|$:  The
positive-helicity reference line gives the reference momentum for negative
helicity, and vice versa.)  There are only three diagrams; however, the +
reference line uses only the ++$-$ vertex, while the $-$ reference line uses
only the $--$+ vertex, so the 4-point-vertex diagram vanishes, as does the
diagram with both reference lines at the same vertex.  Thus, we are left with
only 1 graph.  Furthermore, we know that the 3-point vertices contribute only
1 term to the reference line, so this graph has only 1 term.  This means we
can immediately write down the answer:
$$ A_{\rm tree}(++--)=  
 ·_{2+}{}^+ ·_{3-}{}^- p_2 p_3 {1\over ü(P_3+P_4)^2}   
$$
$$ 
= {[-2]\over Ò+2Ô}{Ò+3Ô\over [-3]}Ò+2Ô[-2]Ò+3Ô[-3]
	{1\over Ò34Ô[34]}{1\over Ò+-Ô[-+]} 
= {[12]^2 Ò34Ô\over [34][41]Ò14Ô}  
\(37) 
$$
 where we have restored helicity and dimensions.  Using the identities,
following from overall momentum conservation,
$$ (P_1+P_4)^2 = (P_2+P_3)^2âÜâ[41]Ò14Ô = [23]Ò32Ô  
\(38) $$
$$ Ý|pÔ[p| =0âÜâÒ34Ô[14] = -Ò32Ô[12]  
\(39) $$
 this can be put in the standard form
$$ A_{\rm tree}(++--)=  {[12]^4\over [12][23][34][41]} 
\(40) $$
 (Similar manipulations cast it into the form $Ò34Ô^4/Ò12ÔÒ23ÔÒ34ÔÒ41Ô$.)

By comparison, the Fermi-Feynman gauge would produce 75 terms for this
calculation, and the Gervais-Neveu gauge 19 terms.  The usual spinor helicity
methods eventually reduce this to the same 1 graph and 1 term, but only
after examining the vector products among polarization vectors and
momenta.  Although relatively simple in this example (the easiest), tracking
indices across diagrams (through propagators) can be tedious in general.  The
spacecone eliminates this index algebra, since the fields are effectively
scalars.  Similar remarks apply to amplitudes involving fermions, where spinor
algebra is involved when spinor helicity is applied to the usual Feynman rules,
but no such algebra appears on the spacecone.  Of course, the final result
will always contain Lorentz products, and some manipulation may be used for
further simplification, but no algebra whatsoever is required on the
spacecone to eliminate vanishing graphs or terms.

A more complicated example is the +++$--$ amplitude.  Again taking
color-ordered (planar) amplitudes, we choose the amplitude cyclically ordered
as +++$--$ with lines labeled 12345, picking 1 and 5 as the reference lines,
which we denote as $¢$++$-\¢$.  Again dropping all graphs with a reference
line at a 4-point vertex or 2 references lines at a 3-point, all 5 graphs with a
4-point vertex are killed, and only 3 of the remaining 5 survive.  Since 3-point
vertices with (without) a reference line have 1 (2) terms, we are left with
only 6 terms.  (We also need to consider various combinations of + and $-$
indices, but only 1 survives for each graph because of the chirality of
3-vertices with reference lines.)  The initial result for the amplitude is then
$$ A_{\rm tree}(+++--)=  
$$ 
$$
·_{2+}·_{3+}·_{4-} \left[ {p_2 p_4^2  \left( 
	{p_2^- +1 \over p_2} -{p_3^- \over p_3} \right) \over (P_1ÉP_2)(P_4ÉP_5)}
	-{p_4^3 \left( 
	{p_2^- \over p_2} -{p_3^- \over p_3} \right) \over (P_2ÉP_3)(P_4ÉP_5)}  
-{p_2^2 p_4 \left( 
	{p_2^- +1 \over p_2} -{p_3^- \over p_3} \right) \over (P_1ÉP_2)(P_3ÉP_4)
} 
\right] 
\(41) $$
where we have used the fact that the reference lines have trivial momenta:
1 for the component with $à$ index opposite to its helicity, 0 for the
remaining components.  The two terms for each diagram simplify to one, using
$$ {p^-\over p} = {[p+]\over [-p]}âÜâ
	{p_2^- \over p_2} -{p_3^- \over p_3} =
	{[2+][-3] -[3+][-2]\over [-2][-3]} = {[23]\over [-2][-3]}  
\(42) $$
 with our normalization.  Using this result, we find the similar result
$$ {p_2^- +1 \over p_2} -{p_3^- \over p_3} =
	{Ò+-Ô[-3] +Ò+2Ô[23] \over Ò+2Ô[-2][-3]} = {Ò+4Ô[34]\over Ò+2Ô[-2][-3]} 
\(43) $$
 applying momentum conservation.    We next translate the momentum
denominators into twistor notation, and also substitute the spacecone
expressions for the polarizations and numerators.  Cancelling identical factors
in numerator and denominator (but no further use of identities), the
amplitude becomes ($+=5$, $-=1$)
$$ {Ò+4Ô^3 \over Ò+2ÔÒ+3Ô}\left( {[-4][34]\over Ò2-Ô[4+]}
	+{[-4]^2 \over Ò23Ô[4+]} +{Ò+2Ô[-2] \over Ò2-ÔÒ34Ô} \right)  
$$
$$ = {Ò+4Ô^3 \over Ò+2ÔÒ+3Ô}\left( -{Ò+2Ô[-4]\over Ò2-ÔÒ23Ô} 
	+{Ò+2Ô[-2] \over Ò2-ÔÒ34Ô} \right)
	= -{Ò+4Ô^3 \over Ò2-ÔÒ23ÔÒ34Ô} 
$$ 
$$
= -{Ò45Ô^4 \over Ò12ÔÒ23ÔÒ34ÔÒ45ÔÒ51Ô} 
\(44) $$
 applying momentum conservation twice, restoring normalization, and
replacing the numerals for $à$.

By comparison, in the usual spinor helicity formalism we find only the same 3
diagrams contribute:  In this formalism also, the two reference momenta
can't be on the same 3-point vertex.  Unlike the spacecone, a reference
momentum can attach to a 4-point vertex as long as the opposite line is
off-shell; fortunately, that isn't possible in this example.  In general, a
3-point vertex with an attached reference line and one other external line can
now have 2 terms instead of 3; the same is true for a 3-point vertex with
two external lines of the same helicity.  This reduces the naive 81 terms to
42.  More detailed algebra eventually reduces this to 8 terms.  The algebra
involved in enumerating these terms is comparable in difficulty to that of
all the rest of the calculation.  (In practice, this amplitude is evaluated by
applying supersymmetry to the 3-gluon-2-quark amplitude.)  

Furthermore, the algebra of the nonvanishing terms is more complicated with
the usual spinor helicity, because vector products have to be evaluated in
terms of spinor products, after which terms can be collected.  For example,
the inner product of two polarization vectors in the usual spinor helicity
formalism is
$$ ·_{i+}É·_{j-} = {Ò+|[i| \over Ò+iÔ} É {|jÔ|-] \over [-j]} =
	-{[-i] \over Ò+iÔ}¼{Ò+jÔ \over [-j]} = -(·_{i+})^+ (·_{j-})^-  
\(45) $$
 in spacecone language; the inner product of two vectors has been replaced
with the product of two scalars.  Similar remarks apply to
$$ ·_{ià,ref} É P_j = Ò+|[-| É |jÔ|j] = Ò+jÔ[-j] = p_j  
\(46) $$
 which is just the transverse momentum component, appearing in the
spacecone 3-point vertex.  With the spacecone method, this algebra has
already been done, which allows some collection of terms and common
factors even before substituting explicit twistor expressions.  In particular,
the polarizations now appear as an overall scalar factor for the whole
amplitude; the rest is all momenta, so identities involving momentum
conservation can be applied more generally between graphs.

Ü6. Recursion relations

Another method used to derive higher-point amplitudes is the classical
Schwinger-Dyson equations, i.e., the classical field equations with
perturbative (multiparticle) boundary conditions at infinite times.  (In the
literature, the field has often been mistaken for the current; as usual, these
are distinguished by the fact the field always has an external propagator,
while the current has it amputated, since $õÄ+...=J$.)  The steps are:  (1)
Calculate the first few terms in the series (enumerated by the number of
external lines).  (2) Guess the general result.  (3) Prove that it is correct by
induction, using the Schwinger-Dyson equations.  Of course, the second part is
the hardest in general (at least when one simplifies the third step by using
spacecone methods), and has been possible for just a couple of cases, only
because the results for those cases are so simple.

The solution to the classical field equations is given by tree graphs with all
external lines but one (the field itself) amputated and put on shell [9].  (The
usual external-line wave functions describe the asymptotic field, which is
free.)  This method has been used to find several tree amplitudes at finite 
point order as well as the two well known 
cases with all the on-shell lines possessing the
same helicity, or one different [4,7].  Note that the field
$a^à$ has a $¦$ associated with the opposite end of its external propagator. 
We then see in the former case, with all +'s on on-shell lines, that $a^-$
vanishes because there are no fully-amputated diagrams, even off-shell, with
only +'s externally (again counting +'s and $-$'s on vertices).  Similarly, for the
latter case, with only one $-$ on an on-shell line, we see that $a^-$ has only
++$-$ vertices; but setting that one on-shell $-$ to be a reference line (which
by definition must be on-shell), it is not allowed such a vertex, so $a^-$
vanishes also in this case.  This constrasts with the usual spinor helicity
method, where the analog of the vanishing of $a^-$ requires an inductive
proof.  By similar reasoning, we see that $a^+$ in the former case consists
entirely of ++$-$ vertices; and in the latter case consists of all ++$-$ except
for one $--$+ (no ++$--$), which must have the $-$ reference line directly
attached.  The appearance of only the self-dual field ($a^+$) and almost only
the self-dual vertex (++$-$) means that in both cases one is 
essentially solving the field equations in the self-dual theory [6].

We now consider in more detail the simpler (former) example (the one which
does not directly give a nontrivial scattering amplitude).  As a slight
simplification, we look at the recursion relation for the field as defined in the
self-dual theory:  From the results at the end of section 2, we write
$$ a^+ = pÄ  
\(47) $$
 (Implicitly, we also have $a^-=p^{-1}öÄ$.  These redefinitions make the ++$-$
vertex local.)  The recursion relation is now
$$ Ä(1,n) = {1\over üP^2(1,n)}Ý_{i=1}^{n-1}Ä(1,i)Ä(i+1,n)
	[p^-(1,i)p(i+1,n) -p(1,i)p^-(i+1,n)]  
\(48) $$
$$ P(j,k) ­ Ý_{m=j}^k P_m  
\(49) $$
 where we again use color ordering, number the external lines cyclically, and
$Ä(j,k)$ denotes the field with on-shell lines with momenta $P_j$ through
$P_k$.  (Thus, on the left-hand side of the equation the field has $n$ on-shell
lines, while on the right-hand side the two fields have $i$ and $n-i$.)  Plugging
in the twistor expressions for the vertex momenta, we find
$$ p^-(1,i)p(i+1,n) -p(1,i)p^-(i+1,n) = Ý_{j=1}^i Ý_{k=i+1}^n Ò+jÔ[jk]Ò+kÔ  
\(50) $$

If we are clever we can guess the general result from explicit evaluation
of the lower-order graphs; instead we find in the literature [7], after the
above redefinition,
$$ Ä(i,j) = {1\over Ò+iÔÒi,i+1ÔòÒj-1,jÔÒ+jÔ} 
\(51) $$
 For the initial-condition case $n=1$ this is simply the statement that the
external line factor for $Ä$ is now
$$ ·_Ä = {(·_+)^+\over p} = {1\over Ò+pÔ^2} 
\(52) $$
 The induction hypothesis is also easy to check:  The product of the two $Ä$'s
from the induction hypothesis gives the desired result by itself up to a
simple factor:
$$ Ä(1,i)Ä(i+1,n) = Ä(1,n){Òi,i+1Ô\over Ò+iÔÒ+,i+1Ô} 
\(53) $$
 (The algebra of the color indices works as usual.)  We then perform the sum
over $i$ before that over $j$ and $k$ (the complete sum is over all $i,j,k$ with
$1²j²i<k²n$), making use of one of the identities used in the usual spinor
helicity evaluation,
$$ {ÒabÔ\over Ò+aÔÒ+bÔ} +{ÒbcÔ\over Ò+bÔÒ+cÔ} = {ÒacÔ\over Ò+aÔÒ+cÔ}âÜâ
	Ý_{i=j}^{k-1}{Òi,i+1Ô\over Ò+iÔÒ+,i+1Ô} = {ÒjkÔ\over Ò+jÔÒ+kÔ} 
\(54) $$
 Multiplying this by the vertex momentum factor gives a sum over $j<k$ of
$ÒjkÔ[jk]=P_jÉP_k$, cancelling the external propagator, yielding the desired
result.

Ü7. Conclusions

We have introduced a complex gauge formalism useful for 
reducing further the algebraic complexity associated with 
vector amplitude calculations.  The results here generalize naturally to
theories containing additional matter.  We have seen that the spacecone
formalism simplifies the algebra in massless Feynman diagrams to such an
extent that supersymmetry identities are no longer needed.  Index algebra at
intermediate stages of the calculation is eliminated; the only remnant of spin
indices on fields is the $à$ label for helicity of the two components of the
spacecone Yang-Mills field.  Our complex spacelike gauge, together with the
use of reference momenta, incorporates all the ideas of spinor helicity, and
extends them to internal lines and vertices through lightcone-type
techniques.  

Although we have considered only tree graphs in this paper, the method
applies straightforwardly to loops.  Alternatively, since auxiliary fields are
often useful in the analysis of effective actions (including renormalization),
one might use a background field formalism, where a Fermi-Feynman or
Gervais-Neveu gauge would be used to calculate the gauge-invariant
effective action first, and then the effective action would be evaluated in the
spacecone gauge to derive S-matrix elements.

ÜACKNOWLEDGMENTS

This work was supported in part by the National Science Foundation
Grant No.¼PHY 9722101.

\refs

£1 G. Chalmers and W. Siegel, hep-ph/9708251 and Simplifying Algebra 
        in Feynman Graphs, Part III: Massive Vectors (in preparation).

£2 S. Weinberg, ÓPhys. Rev.Õ É150 (1966) 1313;\\
	J.B. Kogut and D.E. Soper, \PRD 1 (1970) 2901.

£3 R. Penrose, ÓJ. Math. Phys.Õ É8 (1967) 345, 
	ÓInt. J. Theor. Phys.Õ É1 (1968) 61;\\
	M.A.H. MacCallum and R. Penrose, ÓPhys. Rep.Õ É6C (1973) 241;\\
	A. Ferber, \NP 132 (1978) 55.

£4 P. De Causmaecker, R. Gastmans, W. Troost, and T.T. Wu, 
	\NP 206 (1982) 53;\\
	F.A. Berends, R. Kleiss, P. De Causmaecker, R. Gastmans, W. Troost,
	and T.T. Wu, \NP 206 (1982) 61;\\
	Z. Xu, D.-H. Zhang, and L. Chang, \NP 291 (1987) 392;\\
	J.F. Gunion and Z. Kunszt, \PL 161 (1985) 333;\\
	R. Kleiss and W.J. Sterling, \NP 262 (1985) 235;\\
	M. Mangano and S.J. Parke, ÓPhys. Rep.Õ É200 (1991) 301;\\   
       Z. Bern, D. Kosower, and L. Dixon, hep-ph/9602280, ÓAnn. Rev. Nucl. 
        Part. Sci.Õ É46 (1996) 109.

£5  C.N. Yang, \PR 38 (1977) 1377;\\
	A.N. Leznov, ÓTheor. Math. Phys.Õ É73 (1988) 1233,\\
	A.N. Leznov and M.A. Mukhtarov, ÓJ. Math. Phys.Õ É28 (1987) 2574;\\
	A. Parkes, \PL 286B (1992) 265.

£6 W.A. Bardeen, ÓProg. Theor. Phys. Suppl.Õ É123 (1996) 1;\\
	D. Cangemi, hep-th/9605208, \NP 484 (1997) 521;\\
       G. Chalmers and W. Siegel, hep-th/9606061, \PRD 54 (1996) 7628.

£7 S.J. Parke and T. Taylor, \NP 269 (1986) 410, \PR 56 (1986) 2459;\\
	F.A. Berends and W.T. Giele, \NP 306 (1988) 759;\\
	Z. Bern, G. Chalmers, L. Dixon, and D.A. Kosower, hep-ph/9312333, 
	\PR 72 (1994) 2134;\\
	G.D. Mahlon, hep-ph/9312276, \PRD 49 (1994) 4438.

£8 M.T. Grisaru, H.N. Pendleton, and P. van Nieuwenhuizen, \PR 15
	(1977) 996; \\ 
	M.T. Grisaru and H.N. Pendleton, \NP 124 (1977) 333;\\
	M.T. Grisaru and W. Siegel, \PL 110B (1982) 49.

 £9 D.G. Boulware and L.S. Brown, ÓPhys. Rev.Õ É172 (1968) 1628.

\bye